# CROSS-LAYER LINK ADAPTATION DESIGN FOR RELAY CHANNELS WITH COOPERATIVE ARQ PROTOCOL

*Morteza Mardani, Jalil S. Harsini and Farshad Lahouti*
Emails: m.mardani@ece.ut.ac.ir, j.harsini@ece.ut.ac.ir, lahouti@ut.ac.ir
WMC Lab., School of Electrical and Computer Engineering, University of Tehran

## ABSTRACT

The cooperative automatic repeat request (C-ARQ) is a link layer relaying protocol which exploits the spatial diversity and allows the relay node to retransmit the source data packet to the destination, when the latter is unable to decode the source data correctly. This paper presents a cross-layer link adaptation design for C-ARQ based relay channels in which both source and relay nodes employ adaptive modulation coding and power adaptation at the physical layer. For this scenario, we first derive closed-form expressions for the system spectral efficiency and average power consumption. We then present a low complexity iterative algorithm to find the optimized adaptation solution by maximizing the spectral efficiency subject to a packet loss rate (PLR) and an average power consumption constraint. The results indicate that the proposed adaptation scheme enhances the spectral efficiency noticeably when compared to other adaptive schemes, while guaranteeing the required PLR performance.

## 1. INTRODUCTION

Recently, cooperative relaying has emerged as a powerful spatial diversity technique for improved performance over direct transmission. The relay channel and the associated practical relaying protocols are extensively studied in the literature, e.g., in [1]. The incremental decode-and-forward (DF) and selective DF relaying protocols are presented to achieve a higher spectral efficiency over the relay channel [1]. The cooperative automatic repeat request (C-ARQ) is a link level relaying protocol, which exploits the benefits of both incremental and selective DF relaying protocols [2-3]. This paper presents a cross-layer approach for link adaptation design over a relay channel employing C-ARQ protocol at the data-link layer.

It is well known that link adaptation at the transmitter based on adaptive modulation and coding (AMC) and power control is a powerful technique to enhance the system spectral efficiency (SE) over time-varying fading channels [4-5]. To exploit the benefits of power and rate adaptation over the relay channel, recently, several studies have been reported in the literature, e.g., [6]-[8]. In [6], the authors proposed a discrete rate and power adaptation policy for the DF relay channel, in which the source and relay transmit with the same power level and coding rate. They have shown that for a finite-rate feedback, a higher throughput gain is achieved by using higher coding rates and allowing some outage probability for data transmission. A constant-power adaptive modulation scheme for a relay network taking advantage of incremental amplify-and-forward (AF) relaying protocol is proposed in [7], where the selected relay forwards the overheard data to the destination, only when the source-destination channel is in the outage mode. In [8], the performance of a constant-power AMC scheme in AF cooperative systems is analyzed.

Motivated by the above reported studies, for a relay network employing C-ARQ protocol, devising a cross-layer approach for link adaptation design which takes both channel conditions and packet-level QoS constraints into consideration, is an interesting research problem. This is the focus of the current article. We consider a single-relay transmission system in which both source and relay nodes are equipped with AMC and power adaptation capabilities at the physical layer, and the relay node retransmits the packets erroneously received at the destination. In general, a relay selection algorithm may be employed to select one relay node among a large number of potential relays in the network, so that the probability of correctly decoding the source packets at the selected relay is sufficiently high.

We first derive closed-form expressions for both the system spectral efficiency performance and the system average transmit power for the considered single-relay scenario with C-ARQ protocol. We then optimize both AMC and power adaptation schemes such that the system spectral efficiency is maximized subject to a system average power and target packet loss rate (PLR) constraint at the data-link layer. To this end, a low complexity iterative algorithm is proposed. We also present a constant-power rate-adaptive C-ARQ protocol for the considered relay channel optimizing the spectral efficiency performance. This particularly suits networks with limited channel state information (CSI) feedback and simple transmitter structure.

The results indicate that the proposed link adaptation schemes noticeably enhance the spectral efficiency performance compared to an adaptive-rate direct transmission system, while satisfying the prescribed QoS requirements.

## 2. SYSTEM MODEL

### 2.1. Protocol Description

As illustrated in Fig. 1, we consider a wireless network composed of a source node (S), a relay node (R) and a destination node (D), where each node is equipped with a single antenna. At the S node, input packets from higher layers of protocol stack are first stored in a transmit buffer, grouped into frames, and then transmitted over the wireless channel on a frame by frame basis. We adopt the packet and frame structure as in [9], where the cyclic redundancy check (CRC) bits of each packet facilitate perfect error detection. The time is slotted, so that in each time-slot the S node (resp. R node) transmits a data frame with duration $T_f$ (resp. $\alpha T_f$). The parameter $0<\alpha<1$ may be chosen to satisfy the latency accepted for decoding a packet at the destination. The proposed C-ARQ protocol acts as follows: the S node first transmits a data frame in the current time-slot. Upon successful reception of all packets in this frame at the D node, it broadcasts an ACK message and the S node transmits a new data-frame in the next time-slot. In case, the D node receives a packet in error, it transmits a NACK message identifying the corrupted packet. The R node if it has successfully received the corrupted packet, places it in a transmit buffer, and retransmits it to the D node each time a frame of duration $\alpha T_f$ is constructed. A simple receiver is employed at the D node that drops the corrupted packets and only decodes based on the retransmitted packets.

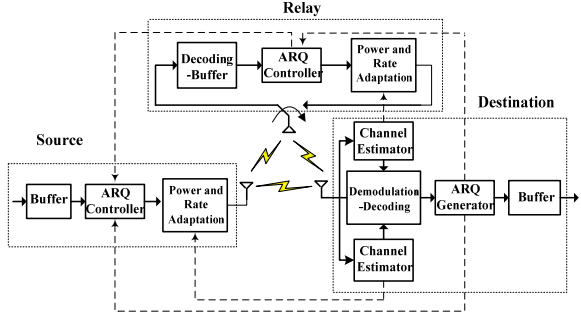

Fig. 1. System model

### 2.2. Channel Model and Physical-Layer Transmission

The S-D, R-D and S-R channels are modeled as discrete time channels with stationary and ergodic time varying power gains $g_{sd}$, $g_{rd}$ and $g_{sr}$, respectively, and AWGN with one-sided power spectral density $N_0$. We adopt a block fading model for the S-D and R-D channels, so that the channel gains remain constant over a frame, but are allowed to vary from one frame to another. This model suits wireless links with slowly-varying fading [9]. Let $W$ denote the bandwidth of the transmitted signal, and $\bar{P}_s$ and $\bar{P}_r$ denote the average transmit powers at the S and R nodes, respectively. We define the pre-adaptation received SNRs of the S-D, R-D and S-R links as $\gamma_1 = \bar{P}_s g_{sd}/N_0 W$, $\gamma_2 = \bar{P}_r g_{rd}/N_0 W$ and $\gamma_3 = \bar{P}_s g_{sr}/N_0 W$, respectively. In each time slot, the S (or R) node adapts its transmit power based on the power control policy $P_s(\gamma_1)$ (or $P_r(\gamma_2)$), thus, the post-adaptation received SNRs of the S-D, R-D and S-R links are $\gamma_1^{ps} = P_s(\gamma_1)\gamma_1/\bar{P}_s$, $\gamma_2^{ps} = P_r(\gamma_2)\gamma_2/\bar{P}_r$ and $\gamma_3^{ps} = P_s(\gamma_1)\gamma_3/\bar{P}_s$, respectively.

The AMC is employed on both S-D and R-D links by dividing the entire SNR range of each link into $N+1$ non-overlapping consecutive intervals. Let $\{\Gamma_{1,n}\}_{n=1}^N$ and $\{\Gamma_{2,m}\}_{m=1}^N$, represent the SNR thresholds for the S-D and R-D links, respectively, where $\Gamma_{i,0} = 0$ and $\Gamma_{i,N+1} = \infty$, $i=1,2$. When the SNR $\gamma_1$ falls in the interval $[\Gamma_{1,n}, \Gamma_{1,n+1})$, the mode $n$ of AMC is chosen and the S node transmits at the rate of $R_n$ bits/symbol and power $P_{s,n}(\gamma_1)$. In the same manner, when the SNR $\gamma_2$ falls in the interval $[\Gamma_{2,m}, \Gamma_{2,m+1})$, the R node transmits at the rate $R_m$ and power $P_{r,m}(\gamma_2)$. No signal is transmitted when $\Gamma_i \in [\Gamma_{i,0}, \Gamma_{i,1})$, $i = 1,2$, corresponding to the outage modes of S-D and R-D links. It is assumed that the SNR estimation at the D node is perfect and that the estimated SNR is fed back to the corresponding transmitter node reliably and without delay.

To facilitate the analysis, we approximate the packet error rate (PER) over each frame corresponding to an AWGN channel with the post adaptation SNR using the following expression [5],

$$PER_{i,n}(\gamma_i) \approx \begin{cases} 1, & \gamma_i < \Gamma_{pn} \\ a_n \exp(-g_n \gamma_i^{ps}), & \gamma_i \geq \Gamma_{pn} \end{cases} \quad (1)$$

where $i = 1,2$ refers to S-D and R-D links, respectively. The parameters $\{a_n, g_n, \Gamma_{pn}\}$ are determined by curve fitting to the exact PER of mode $n$.

### 3. LINK ADAPTATION FOR RELAY CHANNEL

In this section, we develop a cross-layer approach to design the AMC and power control modules over the relay channel when a C-ARQ is employed at the data-link layer. We focus on a specific scenario of interest as described by the following two assumptions:

A1) We assume that the maximum number of retransmission attempts per packet at the R node is limited to one, which guarantees a low packet delay. As demonstrated in [5] for the case of a point-to-point wireless link, a joint adaptive-power AMC-ARQ with only one retransmission almost achieves the maximum possible SE gain over a block-fading channel.

A2) We consider a specific setting with reliable S-R transmission. This is a realistic assumption in wireless networks with a large number of potential relays, so that the S node can select a relay node with a good channel condition [10]. The study of such a relay selection algorithm is beyond the scope of this paper.

The error free delivery of packets, using one retransmission attempt, is not guaranteed. Therefore, if a packet is not received correctly following a possible relay retransmission, it is considered lost. Accordingly, we assume that the packet service to be provisioned imposes a PLR constraint at the data-link layer. Having specified the C-ARQ protocol and the QoS constraints, we next aim at addressing the following interesting question: benefiting from the spatial diversity of relay retransmission which alleviates the stringent error control requirements over the S-D link, how we can design the adaptation scheme at the physical layer to maximize the spectral efficiency while satisfying the packet-level QoS requirements? To this end, we first analyze the spectral efficiency performance of the considered system.

### 3.1. Spectral Efficiency Performance

As in [11], we define the average spectral efficiency as the average number of *accepted* information bits per transmitted symbol. The next Proposition presents an exact closed-form expression for the average SE of the proposed scheme.

*Proposition* 1: The average spectral efficiency of the considered packet based adaptive power and rate C-ARQ scheme is given by

$$\eta = \sum_{n=1}^N R_n \left(1 - \overline{PER}_{1,n}\right) \pi_{1,n}$$
$$+ \sum_{n=1}^N \sum_{m=1}^N \frac{R_n R_m}{R_n + R_m} \overline{PER}_{1,n}(1 - \overline{PER}_{2,m}) \pi_{2,m} \pi_{1,n} \quad (2)$$

where mode $n$ over link $i$ is chosen with the probability $\pi_{i,n} \triangleq \int_{\Gamma_{i,n}}^{\Gamma_{i,n+1}} p_{\Gamma_i}(\gamma) d\gamma$, $i=1,2$, and $p_{\Gamma_i}(.)$ is the probability density function of SNR $\gamma_i$. The average PER in mode $n$ is also given by

$$\overline{PER}_{i,n} \triangleq \frac{1}{\pi_{i,n}} \int_{\Gamma_{i,n}}^{\Gamma_{i,n+1}} PER_{i,n}(\gamma) p_{\Gamma_i}(\gamma) d\gamma$$

*Proof*: The proof is provided in Appendix A.

### 3.2. Optimizing Spectral Efficiency

The objective of the link adaptation scheme is to adjust the AMC mode switching levels and the transmit power levels of the S and R nodes such that the spectral efficiency is maximized subject to a target PLR and an average transmit power constraint as follows

$\max_{\{\Gamma_{1,n},\Gamma_{2,m},P_{s,n}(\gamma_1),P_{r,m}(\gamma_2)\}_{n,m=1}^N} \eta$ subject to

$C_1: P_{avg} \leq \bar{P}$

$C_2: PLR \leq P_{loss}$ (3)

Here, $P_{avg}$ and $\bar{P}$, denote the average transmit power and the maximum average power, respectively, and $C_1$ represents the system average transmit power constraint. Besides, $PLR$ and $P_{loss}$ indicate the system instantaneous PLR and the target PLR, respectively, and $C_2$ is the instantaneous PLR constraint.

If a packet is not received correctly by the D node following the S node transmission, the relay retransmits it. The instantaneous network PLR is thus given by

$$PLR = PER_{1,n}(\gamma_1)PER_{2,m}(\gamma_2) \quad (4)$$

Satisfying the constraint $C_2$ in (3) with equality leads to the equation $PER_{1,n}(\gamma_1) = P_{loss}/PER_{2,m}(\gamma_2)$. Since both sides of this equation are functions of different random variables, we set $PER_{1,n}(\gamma_1) = P_{t,1}$ and $PER_{2,m}(\gamma_2) = P_{t,2}$, where the constants $P_{t,1}$ and $P_{t,2}$ denote target PERs for the S-D and R-D links, respectively. These target PERs must satisfy the equality $P_{t,1}P_{t,2} = P_{loss}$. Accordingly, using (1) the following power adaptation policies for the S and R nodes can be obtained

$$P_{s,n}(\gamma_1)/\bar{P}_s = h_{1,n}/\gamma_1, \quad \Gamma_{1,n+1} > \gamma_1 > \Gamma_{1,n} \geq \Gamma_{pn} \quad (5)$$

$$P_{r,m}(\gamma_2)/\bar{P}_r = h_{2,m}/\gamma_2, \quad \Gamma_{1,m+1} > \gamma_2 > \Gamma_{2,m} \geq \Gamma_{pm} \quad (6)$$

where $h_{1,n} = \frac{1}{g_n}\ln(a_n/P_{t,1})$ and $h_{2,m} = \frac{1}{g_m}\ln(a_m/P_{t,2})$.

In order to solve problem (3), we also need to determine the system average power as indicated by $C_1$. Let $P^j$ be the system consumed power during the time frame $j=1, 2, \ldots$ with duration $T_f^j$. Considering the ergodic process $\{P^j\}$, we can compute the long term system average power, defined as the ratio of the consumed energy over a long time period to the elapsed time, as follows

$$P_{avg} = \lim_{K\to\infty} \sum_{j=1}^K P^j T_f^j / \sum_{j=1}^K T_f^j \quad (7)$$

The following proposition presents a closed-form expression for the average consumed power of the proposed scheme.

*Proposition* 2: The average power of the considered adaptive power and rate C-ARQ protocol is given by

$$P_{avg} = \frac{1}{1+P_{t,1}\Omega(\Gamma_1,\Gamma_2)}\mathbb{E}[P_s(\gamma_1)] + \frac{P_{t,1}\Omega(\Gamma_1,\Gamma_2)}{1+P_{t,1}\Omega(\Gamma_1,\Gamma_2)}\mathbb{E}[P_r(\gamma_2)] \quad (8)$$

where $\mathbb{E}[.]$ is the statistical expectation operator and $\Omega(\Gamma_1,\Gamma_2) = \sum_{n=1}^N\sum_{m=1}^N \frac{R_n}{R_m}\pi_{2,m}\pi_{1,n}$, where the vectors $\Gamma_1 = (\Gamma_{1,1},\ldots,\Gamma_{1,N})$, and $\Gamma_2 = (\Gamma_{2,1},\ldots,\Gamma_{2,N})$ contain the mode switching levels at the source and relay nodes, respectively.

*Proof*: The proof is provided in Appendix B.

Using equations (5), (6), and (8), the desired optimization problem in (3) can be simplified as follows

$\max_{P_{t,1},\Gamma_1,\Gamma_2} (1-P_{t,1})\sum_{n=1}^N R_n\pi_{1,n}$

$\qquad + (P_{t,1}-P_{loss})\sum_{n=1}^N\sum_{m=1}^N \frac{R_nR_m}{R_n+R_m}\pi_{2,m}\pi_{1,n}$

$C_1: \sum_{n=1}^N \int_{\Gamma_{1,n}}^{\Gamma_{1,n+1}} \frac{\bar{P}_s h_{1,n}}{\gamma} p_{\Gamma_1}(\gamma)d\gamma + P_{t,1}\Omega(\Gamma_1,\Gamma_2)$

$\qquad \times \sum_{m=1}^N \int_{\Gamma_{2,m}}^{\Gamma_{2,m+1}} \frac{\bar{P}_r h_{2,m}}{\gamma} p_{\Gamma_2}(\gamma)d\gamma \leq \bar{P}(1+P_{t,1}\Omega(\Gamma_1,\Gamma_2))$

$C_2: \Gamma_{1,n} \geq \Gamma_{pn}, \Gamma_{2,m} \geq \Gamma_{pm}, \quad n,m=1,2,\ldots,N$ (9)

Note that applying the Lagrange method to solve the problem (9) does not directly lead to a tractable solution. In the next section, an iterative solution is presented instead.

### 4. ITERATIVE SOLUTION FOR LINK ADAPTATION

In this section, we present an iterative method to obtain an optimized solution for the problem in (9). The main complexity in solving (9) arises from the term $\Omega(\Gamma_1,\Gamma_2)$ in the power constraint $C_1$. To enable the analysis, at the $i$th iteration we approximate this term with a fixed value determined based on the solution at the previous iteration, i.e., $\varphi^{(i)} = \Omega(\Gamma_1^{(i-1)},\Gamma_2^{(i-1)})$. For notation brevity, we drop the iteration index $i$, in the following analysis. Accordingly, the Lagrangian of the modified optimization problem can be expressed as follows

$L(\lambda,\{\lambda_{1,n},\lambda_{2,n}\}_{n=1}^N,\Gamma_1,\Gamma_2) =$

$\quad (1-P_{t,1})\sum_{n=1}^N R_n \int_{\Gamma_{1,n}}^{\Gamma_{1,n+1}} p_{\Gamma_1}(\gamma)d\gamma + (P_{t,1}-P_{loss})$

$\quad \times \sum_{n=1}^N\sum_{m=1}^N \frac{R_nR_m}{R_n+R_m}\int_{\Gamma_{1,n}}^{\Gamma_{1,n+1}}\int_{\Gamma_{2,m}}^{\Gamma_{2,m+1}} p_{\Gamma_1}(\gamma_1)p_{\Gamma_2}(\gamma_2)d\gamma_1 d\gamma_2$

$\quad -\lambda\sum_{n=1}^N \int_{\Gamma_{1,n}}^{\Gamma_{1,n+1}} \frac{\bar{P}_s h_{1,n}}{\gamma} p_{\Gamma_1}(\gamma)d\gamma - \lambda\varphi P_{t,1}\times$

$\quad \sum_{m=1}^N \int_{\Gamma_{2,m}}^{\Gamma_{2,m+1}} \frac{\bar{P}_r h_{2,m}}{\gamma} p_{\Gamma_2}(\gamma)d\gamma + \sum_{i=1}^2\sum_{n=1}^N \lambda_{i,n}(\Gamma_{i,n}-\Gamma_{pn})$ (10)

where $\lambda,\{\lambda_{i,n}\}_{n=1}^N$, $i=1,2$ are the Lagrange multipliers. Given a fixed target PER $P_{t,1}$, the optimal solution must satisfy the Karush Kauhn Tucker (KKT) conditions as follows [12]

$\sum_{n=1}^N \int_{\Gamma_{1,n}}^{\Gamma_{1,n+1}} \frac{\bar{P}_s h_{1,n}}{\gamma} p_{\Gamma_1}(\gamma)d\gamma + \varphi P_{t,1}$

$\times \sum_{m=1}^N \int_{\Gamma_{2,m}}^{\Gamma_{2,m+1}} \frac{\bar{P}_r h_{2,m}}{\gamma} p_{\Gamma_2}(\gamma)d\gamma = \bar{P}(1+\varphi P_{t,1})$ (11)

$\frac{\partial L(\lambda,\{\lambda_{1,n},\lambda_{2,n}\}_{n=1}^N,\Gamma_1,\Gamma_2)}{\partial \Gamma_{1,n}} = 0, \quad n=1,2,\ldots,N$ (12)

$\frac{\partial L(\lambda,\{\lambda_{1,n},\lambda_{2,n}\}_{n=1}^N,\Gamma_1,\Gamma_2)}{\partial \Gamma_{2,m}} = 0, \quad m=1,2,\ldots,N$ (13)

$\Gamma_{i,n} \geq \Gamma_{pn} \quad i=1,2, \quad n=1,2,\ldots,N$

$\lambda,\lambda_{i,n} \geq 0, i=1,2, \quad n=1,2,\ldots,N$

$\lambda_{i,n}(\Gamma_{i,n}-\Gamma_{pn}) = 0, \quad i=1,2, n=1,2,\ldots,N$ (14)

Using (14) if $\Gamma_{i,n} \geq \Gamma_{pn}, \forall i,n$, then from the slackness conditions [12] we have $\lambda_{i,n} = 0, \forall i,n$. Hence, from (12) we can obtain the mode switching levels of the S-D link as follows

$\Gamma_{1,1} = f_{1,1}(\Gamma_2,\lambda) \triangleq \max\left\{\frac{\lambda\bar{P}_s h_{1,1}/R_1}{1-P_{t,1}+(P_{t,1}-P_{loss})\sum_{m=1}^N \frac{R_m}{R_m+R_1}\pi_{2,m}}, \Gamma_{p1}\right\}$

$$\Gamma_{1,n} = f_{1,n}(\boldsymbol{\Gamma}_2, \lambda) \triangleq$$
$$\max\left\{\frac{\lambda \bar{P}_s(h_{1,n}-h_{1,n-1})/(R_n-R_{n-1})}{1-P_{t,1}+(P_{t,1}-P_{loss})\sum_{m=1}^{N}\frac{R_m^2}{(R_m+R_{m-1})(R_m+R_n)}\pi_{2,m}}, \Gamma_{pn}\right\}, n \geq 2 \quad (15)$$

From (13), we also obtain the mode switching levels of the R-D link as follows

$$\Gamma_{2,1} = f_{2,1}(\boldsymbol{\Gamma}_1, \lambda) \triangleq \max\left\{\frac{\lambda\varphi\bar{P}_r h_{2,1}/R_1}{(1-P_{loss}/P_{t,1})\sum_{n=1}^{N}\frac{R_n}{R_1+R_n}\pi_{1,n}}, \Gamma_{p1}\right\}$$

$$\Gamma_{2,m} = f_{2,m}(\boldsymbol{\Gamma}_1, \lambda)$$
$$\triangleq \max\left\{\frac{\lambda\varphi\bar{P}_r(h_{2,m}-h_{2,m-1})/(R_m-R_{m-1})}{(1-P_{loss}/P_{t,1})\sum_{n=1}^{N}\frac{R_n^2}{(R_n+R_{m-1})(R_n+R_m)}\pi_{1,n}}, \Gamma_{pm}\right\}, m \geq 2 \quad (16)$$

It is worth noting that for the selected AMC modes and the practical range of the target PERs, the sequences $(h_{1,n} - h_{1,n-1})/(R_n - R_{n-1})$ and $(h_{2,m} - h_{2,m-1})/(R_m - R_{m-1})$, $n,m=1,2,\ldots,N$ are increasing in $n$ and $m$, respectively, which in turn lead to $0 < \Gamma_{1,1} < \cdots < \Gamma_{1,N}$ and $0 < \Gamma_{2,1} < \cdots < \Gamma_{2,N}$.

Iterative Algorithm

As presented in (15) and (16), the mode switching levels of the S-D link are functions of those of the R-D link, and vice versa. Motivated by this observation, here, we propose an iterative algorithm to find the optimized solution $\boldsymbol{\Gamma}_1^*, \boldsymbol{\Gamma}_2^*$, given that the target PER $P_{t,1}$ is fixed. Let $I_{max}$ denotes the maximum number of iterations.

Step1) Initialize $\boldsymbol{\Gamma}_1^{(0)}, \boldsymbol{\Gamma}_2^{(0)}$.

Step2) Repeat ($i$ =1: $I_{max}$)

1) Obtain $\lambda^{(i)}$ so that the constraint $C_1$ in (11) is satisfied with equality; update the followings based on (15) and (16)

$\Gamma_{2,m}^{(i)} = f_{2,m}\left(\boldsymbol{\Gamma}_1^{(i-1)}, \lambda^{(i)}\right), m = 1, \ldots, N$,
$\Gamma_{1,n}^{(i)} = f_{1,n}\left(\boldsymbol{\Gamma}_2^{(i-1)}, \lambda^{(i)}\right), n = 1, \ldots, N$, and
$\varphi^{(i)} = \Omega(\boldsymbol{\Gamma}_1^{(i-1)}, \boldsymbol{\Gamma}_2^{(i-1)})$.

2) Obtain the SE, $\eta^{(i)}$ from (2).

3) Repeat 1 to 3 until $\eta^{(i)}$ converges to $\eta(P_{t,1})$.

The maximum SE corresponds to an optimized target PER,

$$P_{t,1}^* = \operatorname{argmax}_{P_{loss} < P_{t,1} < 1} \eta(P_{t,1})$$

Based on extensive numerical experiments, we observed that the SE is always a quasiconcave function [12] of $P_{t,1}$ in the entire range of the target PER, i.e., $P_{loss} < P_{t,1} < 1$ (see Fig. 3). Therefore, it is straight forward to devise a fast one dimensional numerical search method to find the optimized target PER, $P_{t,1}^*$, (A similar case is presented in [12]). Accordingly, the optimized mode switching levels $\boldsymbol{\Gamma}_1^*, \boldsymbol{\Gamma}_2^*$ are obtained from the converged solution corresponding to the optimized target PER $P_{t,1}^*$.

## 5. ADAPTIVE RATE C-ARQ SCHEME

In this section, we consider a scenario where only the quantized CSI is fed back to the S and R nodes. Specifically, here the CSI refers to the index of the AMC modes to be chosen at the S and R nodes. In this case, we propose an adaptive rate C-ARQ scheme so that the S and R nodes transmit with constant power levels $\bar{P}_s$ and $\bar{P}_r$, respectively. This in turn result in $\gamma_1^{ps} = \gamma_1, \gamma_2^{ps} = \gamma_2$.

Our objective here is to find the optimized mode switching levels of S-D and R-D links that maximizes the spectral efficiency subject to an instantaneous PLR constraint. The SE of a joint constant power AMC with C-ARQ scheme can be obtained by substituting $\gamma_1^{ps} = \gamma_1$, $\gamma_2^{ps} = \gamma_2$ in equation (2). In order to guarantee the instantaneous PLR constraint $PLR \leq P_{loss}$, we set the AMC thresholds for modes $n$, $m$ to the minimum SNRs required to achieve the PLR $P_{loss}$, i.e.,

$$PER_{1,n}(\Gamma_{1,n})PER_{2,m}(\Gamma_{2,m}) = P_{loss} \quad (17)$$

Since both terms in the L.H.S. of (17) are functions of different variables, we set $PER_{1,n}(\Gamma_{1,n}) = P_{t,1}$ and $PER_{2,m}(\Gamma_{2,m}) = P_{t,2}$, where $P_{t,1}$ and $P_{t,2}$ denote the target PER imposed on the S-D and R-D links, respectively.

Given a fixed target PER $P_{t,1}$, using the equation $PER_{1,n}(\Gamma_{1,n}) = P_{t,1}$, the mode switching levels of S-D link are obtained as

$$\Gamma_{1,n} = \max\left\{\frac{1}{g_n}\ln(a_n/P_{t,1}), \Gamma_{pn}\right\}, n = 1,2,\ldots,N \quad (18)$$

From (17), we obtain the target PER $P_{t,2} = P_{loss}/P_{t,1}$. Therefore, the mode switching levels of the R-D link are determined from the equation $PER_{2,m}(\Gamma_{2,m}) = P_{loss}/P_{t,1}$, as follows

$$\Gamma_{2,m} = \max\left\{\frac{1}{g_m}\ln(a_m P_{t,1}/P_{loss}), \Gamma_{pm}\right\}, m = 1,2,\ldots,N \quad (19)$$

As evident from (18) and (19), both sets of mode switching levels $\{\Gamma_{1,n}\}_{n=1}^{N}$ and $\{\Gamma_{2,m}\}_{m=1}^{N}$ depend only on the parameter $P_{t,1}$. Therefore, the optimized mode switching levels $\{\Gamma_{1,n}^*\}_{n=1}^{N}$ and $\{\Gamma_{2,m}^*\}_{m=1}^{N}$ can be obtained from the target PER $P_{t,1}^*$ which is the solution of the following optimization problem

$$P_{t,1}^* = \operatorname{argmax}_{P_{loss} < P_{t,1} < 1} \eta(P_{t,1}) \quad (20)$$

which is solved as discussed in previous section.

## 6. NUMERICAL RESULTS

In this section, we provide numerical results to assess the performance of the proposed schemes. For both the S and R nodes the AMC modes of HYPERLAN/2 standard with a packet length $N_b$ =1080 bits are employed. Table II of [9] presents the AMC transmission modes and their corresponding fitting parameters. We consider a scenario that the channel SNRs $\gamma_1$ and $\gamma_2$ are exponential random variables (Rayleigh fading) with mean $\bar{\gamma}_1 = \bar{P}_s$ and $\bar{\gamma}_2 = \mu\bar{P}_r$, respectively. The parameter $\mu$ represents the effect of path loss on the system performance. We also set $\bar{P}_s = \bar{P}_r = \bar{P}$, and the target PLR to $P_{loss}$=0.001.

Table I depicts the convergence of the proposed iterative algorithm. Our observations validate that using different random initial conditions for mode switching levels, the spectral efficiency increases at each step of the algorithm. Furthermore, after a few iterations, the algorithm converges to a fixed SE value.

Fig. 2 compares the spectral efficiency performance of the proposed joint C-ARQ and AMC scheme with/without power adaptation. In this figure, we also depict the result of an AMC based direct transmission (DT) scheme proposed in [9]. As evident the SE of the proposed power adaptive scheme exceeds that of the constant transmit power scheme by about 10%. Moreover, thanks to the use of relay retransmission, the proposed constant power adaptive-rate C-ARQ scheme attains a 30% SE improvement over to the AMC-based direct transmission system.

Table I
Convergence of iterative algorithm for some random initializations. The SE vs. iterations for $\bar{P} = 10$dB, $\mu = 0$dB.

| SE/Index | 0 | 1 | 2 | 3 | 4 |
|---|---|---|---|---|---|
| Trial #1 | 0.668 | 1.889 | 1.910 | 1.911 | 1.911 |
| Trial #2 | 0.559 | 1.888 | 1.910 | 1.911 | 1.911 |
| Trial #3 | 0.481 | 1.896 | 1.910 | 1.911 | 1.911 |
| Trial #4 | 0.471 | 1.884 | 1.907 | 1.911 | 1.911 |

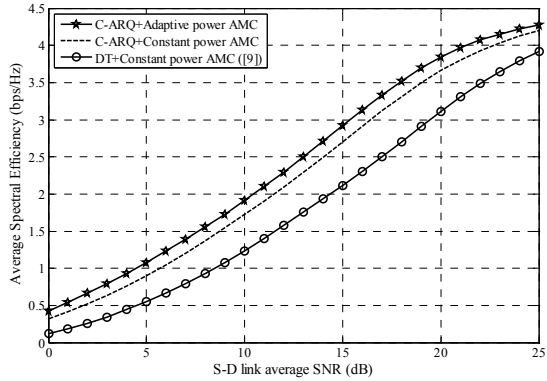

Fig. 2. Spectral efficiency of joint C-ARQ and AMC with/without power control and direct transmission scheme with constant-power AMC, $\mu$=0 dB.

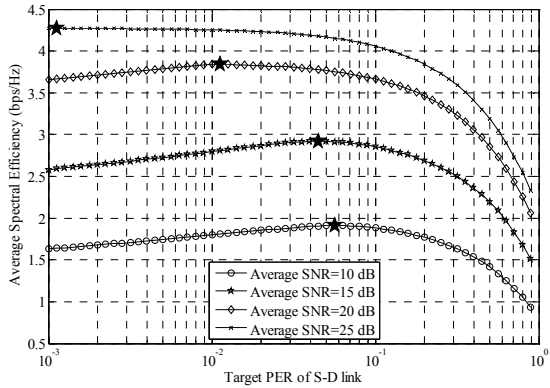

Fig. 3. Spectral efficiency of the proposed scheme vs. target PER of the S-D link, $\mu$=0 dB.

As shown in Fig. 3, the SE is a quasiconcave function in the entire range of target PER of S-D link. Therefore, the search algorithm always finds the optimized solution. From this figure we can also explain how the optimization algorithm in section 4 assigns the target PERs of the S-D and R-D links to optimize the spectral efficiency. As the S-D link quality improves, the corresponding target PER decreases, meaning that this link is able to satisfy the imposed PLR QoS constraint without the relay node contribution. On the other hand, for smaller S-D link channel SNRs, the optimization algorithm relaxes the target PER of the S-D link and relies more on the R-D link retransmission to satisfy the PLR constraint.

## 7. CONCLUSIONS

In this paper, we considered a relay transmission system employing cooperative-ARQ protocol at the data-link layer and AMC and power control at the physical layer. We presented power and rate adaptation schemes which maximize the spectral efficiency subject to a target packet-loss rate and an average system power constraint. Using the Lagrange method, we developed an iterative algorithm to find the optimized link adaptation solution. Numerical results indicate that noticeable spectral efficiency gains are achieved by the proposed adaptation schemes when compared to a constant-power AMC based direct transmission system. Currently, we are investigating the impact of an unreliable S-R link and also a relay selection algorithm on the proposed analysis.

## APPENDIX A

Here, we present a proof for Proposition 1. Let us assume that each $N_b$ bit packet is transmitted using $L$ modulation symbols. The random variable $L$ depends on AWGNs and the channel SNRs in both S-D and R-D links. We define the instantaneous spectral efficiency as the number of accepted information bits per transmitted symbol which is $N_b/L$, if a packet is successfully received at the D node, and otherwise, it is zero.

For the proposed adaptive transmission C-ARQ scheme, each packet may encounter a vector of channel SNRs $\boldsymbol{\gamma} = (\gamma_1, \gamma_2)$. Let the random variables $R_n$ and $R_m$ denote the selected rates by the S and R nodes when the channel SNRs $\gamma_1$ and $\gamma_2$, fall into the intervals $[\Gamma_{1,n}, \Gamma_{1,n+1})$ and $[\Gamma_{2,m}, \Gamma_{2,m+1})$, respectively. Thus, the number of symbols transmitted per packet by the S and R nodes will be $N_b/R_n$ and $N_b/R_m$, respectively. Accordingly, the instantaneous spectral efficiency can be expressed as

$$\eta(\boldsymbol{\gamma}, M) = \begin{cases} \frac{N_b}{N_b/R_n}, & M=1 \\ \frac{N_b}{N_b/R_n + N_b/R_m}, & M=2 \\ 0, & M=3 \end{cases} \quad (21)$$

where the event $M$ is described as

$$M = \begin{cases} 1, & (T_1:s) \\ 2, & (T_1:f) \text{AND} (T_2:s) \\ 3, & (T_1:f) \text{AND} (T_2:f) \end{cases} \quad (22)$$

Here $T_1$ and $T_2$ are the events indicating the success ($s$) or failure ($f$) of the packet transmission over the S-D channel and R-D channels, respectively. These events take the probabilities

$$\Pr(T_1:f) = PER_{1,n}(\gamma_1)$$
$$\Pr(T_1:s) = 1 - \Pr(T_1:f)$$
$$\Pr(T_2:f) = PER_{2,m}(\gamma_2)$$
$$\Pr(T_2:f) = 1 - \Pr(T_2:f)$$

The average spectral efficiency of the proposed scheme, denoted by $\eta$, is obtained by applying the expectation operator to the instantaneous spectral efficiency in (21)

$$\eta = \mathbb{E}_{\boldsymbol{\gamma}} \mathbb{E}_M[\eta(\boldsymbol{\gamma}, M)|\boldsymbol{\gamma}] \quad (23)$$

Averaging with respect to the random variable $M$, the inner expectation in (23) is reduced as

$$\eta(\boldsymbol{\gamma}) = \mathbb{E}_M[\eta(\boldsymbol{\gamma}, M)|\boldsymbol{\gamma}]$$
$$= R_n \left(1 - PER_{1,n}(\gamma_1)\right) + \frac{R_n R_m}{R_n + R_m} PER_{1,n}(\gamma_1)(1 - PER_{2,m}(\gamma_2)) \quad (24)$$

Now, we apply the expectation operator to (24) with respect to the channel SNR realizations, to obtain the average spectral efficiency

$$\eta = \mathbb{E}_{\boldsymbol{\gamma}}[\eta(\boldsymbol{\gamma})]$$
$$= \int_0^\infty \int_0^\infty \left[ R_n \left(1 - PER_{1,n}(\gamma_1)\right) + \frac{R_n R_m}{R_n + R_m} PER_{1,n}(\gamma_1)(1 - PER_{2,m}(\gamma_2)) \right] p_{\Gamma_1}(\gamma_1) p_{\Gamma_2}(\gamma_2) d\gamma_1 d\gamma_2$$
$$= \sum_{n=1}^N R_n \int_{\Gamma_{1,n}}^{\Gamma_{1,n+1}} \left(1 - PER_{1,n}(\gamma_1)\right) p_{\Gamma_1}(\gamma_1) d\gamma_1$$
$$+ \sum_{n=1}^N \sum_{m=1}^N \frac{R_n R_m}{R_n + R_m} \int_{\Gamma_{1,n}}^{\Gamma_{1,n+1}} PER_{1,n}(\gamma_1) p_{\Gamma_1}(\gamma_1) d\gamma_1$$
$$\times \int_{\Gamma_{2,m}}^{\Gamma_{2,m+1}} (1 - PER_{2,m}(\gamma_2)) p_{\Gamma_2}(\gamma_2) d\gamma_2 \quad (25)$$

By defining $\pi_{i,n} \triangleq \int_{\Gamma_{i,n}}^{\Gamma_{i,n+1}} p_{\Gamma_i}(\gamma) d\gamma$, $i=1,2$ and $\overline{PER}_{i,n} \triangleq \frac{1}{\pi_{i,n}}$

$\times \int_{\Gamma_{i,n}}^{\Gamma_{i,n+1}} PER_{i,n}(\gamma) p_{\Gamma_i}(\gamma) d\gamma$, and after following some manipulations, we obtain

$$\eta = \sum_{n=1}^N R_n \left(1 - \overline{PER}_{1,n}\right) \pi_{1,n}$$
$$+ \sum_{n=1}^N \sum_{m=1}^N \frac{R_n R_m}{R_n + R_m} \overline{PER}_{1,n} (1 - \overline{PER}_{2,m}) \pi_{2,m} \pi_{1,n}$$

∎

## APPENDIX B

Here, we present a proof for Proposition 2. Let us consider packet transmission over $K$ consecutive time-slots. We assume that each of S and R nodes transmit at $I$ and $J$ time-slots during this time, respectively (i.e., $I+J=K$). As stated in section 2.2, each of S and R nodes transmit their information with the power $P_s^i$ and $P_r^j$ in frames with duration $T_f$ and $\alpha T_f$, respectively. Therefore, the equation (7) can be written as

$$P_{avg} = \lim_{I,J \to \infty} \frac{\sum_{i=1}^I P_s^i T_f + \sum_{j=1}^J P_r^j \alpha T_f}{\sum_{i=1}^I T_f + \sum_{j=1}^J \alpha T_f}$$

$$= \lim_{I,J \to \infty} \frac{\sum_{i=1}^I P_s^i T_f}{\sum_{i=1}^I T_f + \sum_{j=1}^J \alpha T_f} + \lim_{I,J \to \infty} \frac{\sum_{j=1}^J P_r^j \alpha T_f}{\sum_{i=1}^I T_f + \sum_{j=1}^J \alpha T_f}$$

$$= \lim_{I,J \to \infty} \frac{IT_f}{IT_f + J\alpha T_f} \times \frac{\sum_{i=1}^I P_s^i T_f}{IT_f} + \lim_{I,J \to \infty} \frac{J\alpha T_f}{IT_f + J\alpha T_f} \times \frac{\sum_{j=1}^J P_r^j \alpha T_f}{J\alpha T_f}$$

$$= \lim_{I,J \to \infty} \frac{1}{1+\alpha J/I} \times \lim_{I \to \infty} \frac{\sum_{i=1}^I P_s^i}{I} + \lim_{I,J \to \infty} \frac{\alpha J/I}{1+\alpha J/I} \times \lim_{J \to \infty} \frac{\sum_{j=1}^J P_r^j}{J} \quad (26)$$

Let $\overline{N}_P^f$ denote the average number of packets received in error at the D node. We have

$$\overline{N}_P^f = C \sum_{n=1}^N R_n \overline{PER}_{1,n} \pi_{1,n} \times I \quad (27)$$

where $C \triangleq N_s/N_b$, and the parameter $N_s$ denotes the number of symbols per frame. Accordingly, the average number of transmitted frames by the R node is given by

$$J = \overline{N}_P^f \times N_b \sum_{m=1}^N \pi_{2,m}/R_m \times T_s/(\alpha T_f) \quad (28)$$

Substituting (27) into (28), yields

$$J/I = \frac{1}{\alpha} \sum_{n=1}^N \sum_{m=1}^N \frac{R_n}{R_m} \overline{PER}_{1,n} \pi_{2,m} \pi_{1,n} \quad (29)$$

On the other hand, due to the ergodicity assumption on the power consumption processes at the source and relay nodes, we obtain $\mathbb{E}[P_s(\gamma_1)] = \lim_{I \to \infty} \sum_{i=1}^I P_s^i/I$ and $\mathbb{E}[P_r(\gamma_2)] = \lim_{J \to \infty} \sum_{j=1}^J P_r^j/J$. Using these relations and substituting (29) into (26), we can obtain the average system power as follows

$$P_{avg} = \frac{1}{1+P_{t,1}\sum_{n=1}^N \sum_{m=1}^N \frac{R_n}{R_m}\pi_{2,m}\pi_{1,n}} \mathbb{E}[P_s(\gamma_1)]$$

$$+ \frac{P_{t,1}\sum_{n=1}^N \sum_{m=1}^N \frac{R_n}{R_m}\pi_{2,m}\pi_{1,n}}{1+P_{t,1}\sum_{n=1}^N \sum_{m=1}^N \frac{R_n}{R_m}\pi_{2,m}\pi_{1,n}} \mathbb{E}[P_r(\gamma_2)] \quad (30)$$

∎